\documentclass[12pt]{iopart}

\usepackage{graphicx}
\usepackage{bm}

\usepackage[utf8]{inputenc}
\usepackage[T1]{fontenc}
\usepackage{mathptmx}
\usepackage{etoolbox}
\usepackage{gensymb}
\usepackage{subcaption}
\usepackage[normalem]{ulem}
\usepackage[none]{hyphenat}

\begin{document}

\title[Temperature-dependent investigation of polarisation doping in 330 nm ultraviolet light-emitting diodes]{Temperature-dependent investigation of polarisation doping in 330 nm ultraviolet light-emitting diodes}

\author{P Milner$^{1,2,*}$, V Z Zubialevich$^1$, S M Singh$^{1,2}$\footnote{Present address: Centre for Nanoscience and Technology, University of Sheffield, Western Bank, Sheffield, S10 2TN, UK.}, P Pampili$^1$, B Corbett$^1$ and P J Parbrook$^{1,2,*}$}

\address{$^1$ Tyndall National Institute, University College Cork, Lee Maltings, Dyke Parade, Cork, Ireland}
\address{$^2$ Department of Electrical and Electronic Engineering, University College Cork, Cork, Ireland}
\address{Email: peter.milner@tyndall.ie and p.parbrook@ucc.ie}
\vspace{10pt}

Polarisation doping of Al$_x$Ga$_{1-x}$N, through grading of $x$, has realised major improvements in \emph{p}-type conductivity in ultraviolet (UV) light-emitting diodes (LEDs) compared to conventional impurity doping. However, the exact balance between the two doping regimes to achieve the best device performance is not clear, especially as a function of operating wavelength. In this work, 330 nm LEDs with varied \emph{p}-doping approaches were characterised as a function of temperature: Mg doped only (reference); polarisation doped and Mg doped (co-doped); and polarisation doped only. At room temperature, the co-doped LED showed the highest electroluminescence (EL) intensity, with a similar operating voltage to the reference LED. The highest hole concentration, confirmed by Hall effect measurements, as well as improved injection efficiency revealed by simulations, are credited as the main reasons for EL improvement. A parasitic near-UV luminescence tail, analogous to the "blue luminescence" in \emph{p}-GaN, was observed in both the reference and co-doped LEDs, but was absent in the polarisation doped LED. The reference LED demonstrated the highest increase in operating voltage with decreasing temperature, while the LED with only polarisation doping showed a temperature-independent behaviour, demonstrating the benefits of a polarisation field-induced carrier concentration. Further optimisation of the compositional grading with concurrent Mg doping can potentially produce higher performance LEDs with cleaner spectra. 

\vspace{2pc}

%
%
%

\section{Introduction}

AlGaN-based ultraviolet (UV) light-emitting diodes (LEDs) and laser diodes (LDs) have shown major performance improvements in recent years as a result of technological advancements in crystal quality, increased light extraction and enhanced \emph{p}-type conductivity \cite{kneissl2019, amano2020}. Specifically \emph{p}-type doping of (Al)GaN has been a major bottleneck in the development of UV LEDs and LDs, owing to the large activation energy of Mg, the dominant \emph{p}-type dopant \cite{imura2007, li2002}. Polarisation doping, the technique of engineering the intrinsic polarisation fields in III-Nitrides, has shown particular success in advancing this area through enhanced hole concentrations and mobilities \cite{zhang2022, cao2023, kolbe2023}. 

Linearly grading the Al composition produces a three dimensional polarisation charge, which results in a corresponding three-dimensional equilibrium carrier gas of opposite sign to preserve charge neutrality. The polarisation charge concentration, $\rho_{\pi}$, in a layer of Al$_x$Ga$_{1-x}$N with a linear compositional grade from $x_1$ to $x_2$ over a distance $d$ can be approximated by \cite{simon2010}:
\begin{equation} \label{eq:pol_conc}
\rho_{\pi} = 5\times 10^{13}\left(\frac{x_2-x_1}{d}\right) \mathrm{cm}^{-3}.
\end{equation}

Sato \emph{et al.} \cite{sato2021} reported that for layers with higher Al compositions, it is beneficial to rely purely on polarisation doping, without impurity doping, due to the high concentration of compensating defects introduced by the Mg dopant. For lower Al compositions, in which the formation energies of compensating point defects are higher \cite{van1999doping}, it remains unclear whether concurrent Mg doping is beneficial to polarisation doped devices. 

Furthermore, there presently exists little work exploring this mechanism in LED structures, particularly as a function of temperature. Whereas hole ionisation from impurity atoms is predominantly a thermally activated process, the field-ionised hole concentration is temperature independent. This warrants studying the temperature-dependent device characteristics to give insights into the device physics with polarisation doping. Ishii \emph{et al.} \cite{ishii2020temperature} investigated the  electroluminescence of a polarisation doped 265 nm UV LED grown on a low dislocation density bulk AlN substrate down to 6 K, concluding it is the injection efficiency that is the limiting factor of device performance at room temperature in DUV LEDs. Other groups have reported various other phenomena at low temperatures in conventionally doped UV LED electroluminescence (EL) measurements. In an early work, Chitnis \emph{et al.} \cite{chitnis2002low} reported an increase in EL intensity down to 10 K, which the authors attributed to deep-level states that provide tunnelling transport through the \emph{p}-type layer. Grzanka \emph{et al.} \cite{grzanka2007role} observed quenching of the quantum well (QW) EL below 200 K, while photoluminescence (PL) monotonically increased down to 10 K. However, in a structure without the electron blocking layer (EBL), despite more than two orders of magnitude lower QW intensity, the EL monotonically increased down to 10 K (along with a stronger emission from \emph{p}-type material). Based on these observations, the authors concluded that the hole injection is suppressed at lower temperatures by the valence band offset associated with the EBL. Conversely, Zhang \emph{et al.} \cite{zhang2010low} reported a complete quenching of the QW EL at low temperatures, which was prevented by \emph{inserting} a thin AlN EBL. Clearly, various manifestations of low hole concentrations, either through electron overflow or suppressed hole injection, appear consistently across these conclusions, suggesting it to be the dominant factor affecting electroluminescence behaviour at low temperatures. 

In this work, the goal is to gain an understanding of the potential benefits of polarisation doping, both with and without concurrent Mg doping, to device performance. We have investigated three different \emph{p}-doping regimes in UV LEDs emitting around 330 nm, and explored their electrical and optical behaviour to cryogenic temperatures. The results reveal the importance of optimising doping conditions (through impurity and polarisation doping) to maximise hole injection while minimising parasitic luminescence.

\section{Method}

\begin{figure*}
\centering
\includegraphics[scale=0.55]{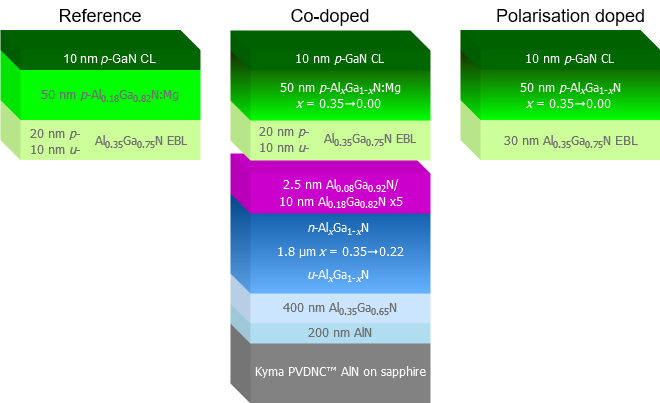}
\caption{\label{fig:fig1}Schematic showing the three LED structures, which are identical except for variations in \emph{p}-doping regimes.}
\end{figure*}

Three LED structures were grown using metalorganic chemical vapour deposition in an Aixtron showerhead-type 3$\times$2" reactor. Trimethylgallium, trimethylaluminium and ammonia were used as the precursors, with disilane and biscyclopentadienyl magnesium (Cp2Mg) used for \emph{n}- and \emph{p}-dopants, respectively. The epitaxial structures are shown schematically in Figure \ref{fig:fig1}. The structure of the three LEDs was nominally identical up to the region between the active region and the top \emph{p}-GaN layer. This consisted of a 200 nm AlN connecting layer on a 2-inch sapphire substrate with a commercially provided sputtered AlN layer. This was followed by 400 nm \emph{u}-Al$_{0.35}$Ga$_{0.65}$N. Next, 1.8 $\mu$m of Al$_x$Ga$_{1-x}$N ($x = 0.35 - 0.22$) was then grown with the upper half \emph{n}-doped to limit the doping-induced tensile strain. The buffer layers show a threading dislocation density (TDD) of approximately 3 $\times 10^9$ cm$^{-3}$ estimated from full width at half maximum of x-ray diffraction (XRD) (002) and (101) measurements. The active region consisted of five 2.5 nm Al$_{0.08}$Ga$_{0.92}$N QWs with 10 nm Al$_{0.18}$Ga$_{0.82}$N barriers, at a growth temperature of 1160$\degree$ C. 

The first (reference) LED, then had a 30 nm Al$_{0.35}$Ga$_{0.65}$N EBL with the upper two thirds Mg-doped. This was followed by 50 nm of Mg-doped \emph{p}-Al$_{0.18}$Ga$_{0.82}$N. The second LED, labelled `co-doped', possesses the same EBL as the reference LED but is followed by 50 nm Mg-doped \emph{p}-Al$_x$Ga$_{1-x}$N ($x = 0.35 - 0.00$). The final `polarisation doped' LED has the same compositional structure as the co-doped LED, except no Cp2Mg was supplied during EBL and \emph{p}-Al$_x$Ga$_{1-x}$N growth. The compositional grading in these \emph{p}-Al$_x$Ga$_{1-x}$N layers forms the negative (for metal-polar growth) polarisation charge concentration, which, according to equation \ref{eq:pol_conc}, corresponds to approximately $3\times10^{18}$ cm$^{-3}$ of negative charges. Secondary ion mass spectrometry (SIMS) measurements (not shown) of a similar structure confirmed the targeted Al compositions in this grading layer. The Mg doping concentration in the \emph{p}-doped layers is approximately $5\times 10^{18}$ cm$^{-3}$. All three structures were capped with 10 nm \emph{p}-GaN:Mg to minimise the contact resistivity, which were activated \emph{in-situ} at 790$\degree$ C in a nitrogen ambient for 20 minutes.  

After growth, the samples were prepared for \emph{p}-metal deposition. This involved a treatment consisting of buffered oxide etchant, dilute hydrochloric acid and a short dip in 45\% potassium hydroxide at 100°C. Soon after, 30 nm Pd was evaporated, which was then annealed at 500°C in a nitrogen ambient. Next, an inductively-coupled Cl$_2$-based plasma etch was used to define the 100 $\mu$m diameter circular mesas and a stack of Ti/Al/Ti/Au was then evaporated for the \emph{n}-metal. SiO$_2$ passivation and Ti/Au bondpad deposition followed to complete the fabrication process.

Initial room temperature characterisation of the LEDs was carried out with an Ocean Insight HR4000 spectrometer and a Keithley 2400 sourcemeter. In each case, a number of individual devices were measured so as to identify a typical behaviour. For the temperature-dependent characterisation, further packaging of typical LEDs was carried out to allow for appropriate mounting. The LEDs were laser diced into smaller arrays and placed on aluminium nitride ceramic submounts. After wirebonding to individual devices and soldering of connections, the samples could be mounted to a closed cycle helium cryostat coldfinger with a thermal paste to ensure good thermal conductivity. Feedthrough wires in the cryostat allowed for the electrical characterisation with the Keithley 2400. Detection of the LED emission was carried out through the sapphire substrate and a small aperture in the ceramic submount using a Horiba iHR320 spectrometer with a Synapse thermoelectrically-cooled CCD detector. Pulsed current injection measurements (not shown) were performed at low temperatures. This confirmed that there was no observable redshift in emission wavelength, which may be indicative of Joule heating with increased duty cycle.

In addition, supplemental test structures related to the \emph{p}-doping strategies in the LEDs were also grown to allow assessment using a LakeShore 8404 Hall measurement system, and to give simpler structures for XRD analysis. These structures, also grown on commercial sputtered AlN-on-sapphire templates, consisted of 1 $\mu$m \emph{u}-Al$_{0.35}$Ga$_{0.65}$N, followed by 100 nm of \emph{p}-Al$_x$Ga$_{1-x}$N (corresponding to the compositional and doping variations in the LED approaches) and were capped with a 20 nm \emph{p}$^+$-GaN contact layer. All structures were annealed for Mg activation under the same conditions as for the LED wafers.

\section{Results and Discussion}

\subsection{Analysis of test structures}
XRD reciprocal space maps of the test structures revealed the top \emph{p}-AlGaN layers to be almost completely strained to the \emph{u}-Al$_{0.35}$Ga$_{0.65}$N layer underneath. Recent work by Yasuda \emph{et al.} has shown the strain preservation to be vital to the conservation of the polarisation field needed for the doping mechanism \cite{yasuda2017relationship}. From our calculations, the fractional relaxation ($\sim$4\%) during the growth of this layer has negligible impact on the desired polarisation field, ensuring the \emph{p}-type conductivity is expected to be maintained.

Initial room temperature Hall measurements carried out on these test structures show that while all three doping approaches led to a free hole concentration, the highest hole concentration was achieved using the co-doped strategy ($p = 6.9\times10^{17}$ cm$^{-3}$). The polarisation only doping approach, although showing good \emph{p}-type conductivity, showed the lowest hole concentration of the three ($p = 2.6\times10^{17}$ cm$^{-3}$).

\subsection{Room temperature characterisation and simulations}
\begin{figure}[htb!]
\centering
\includegraphics[scale=0.5]{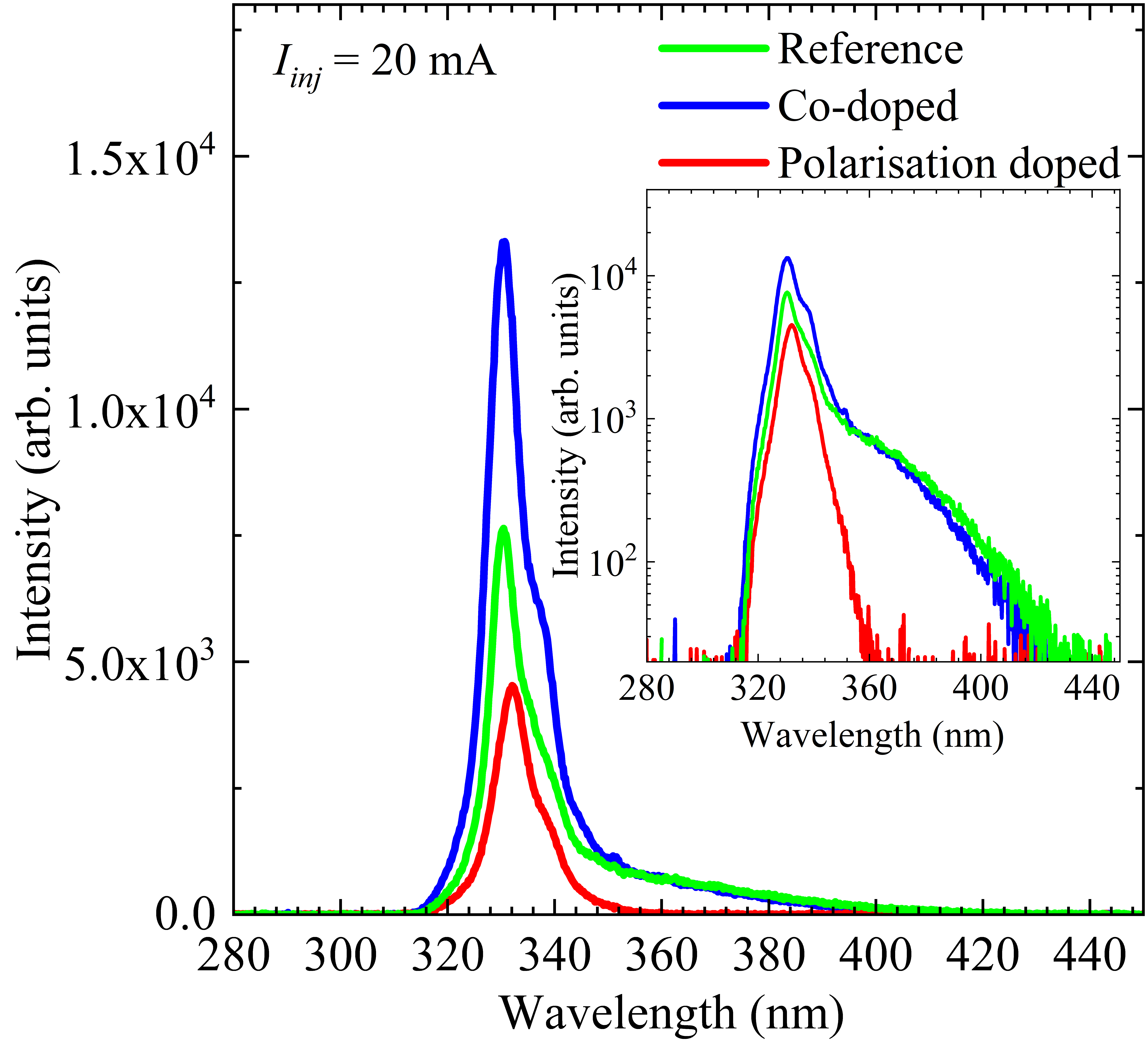}
\caption{\label{fig:fig2} Room temperature EL spectra of the three LEDs. The co-doped LED shows the strongest intensity for the same injection current. Inset are the spectra on a semi-logarithmic scale highlighting the marked reduction of the long wavelength tail in the polarisation doped LED.}
\end{figure}

Figure \ref{fig:fig2} shows the room temperature EL spectra of the three LEDs, measured at 20 mA (approx. 250 A cm$^{-2}$). All LEDs show a distinct QW-related luminescence peak at 330\,nm. When plotted on a semi-logarithmic scale (inset in Figure \ref{fig:fig2}), it is evident that near-UV luminescence (NUVL) centred around 360 nm is present in both of the LEDs with Mg-doped AlGaN, but is absent in the LED without. Based on this observation, we conclude that this NUVL band is a Mg-related defect peak similar to the blue luminescence in \emph{p}-GaN \cite{kaufmann1998}.

\begin{figure}[hbt]
\centering
    \begin{subfigure}[b]{0.5\textwidth}
    \centering
    \includegraphics[scale=0.45]{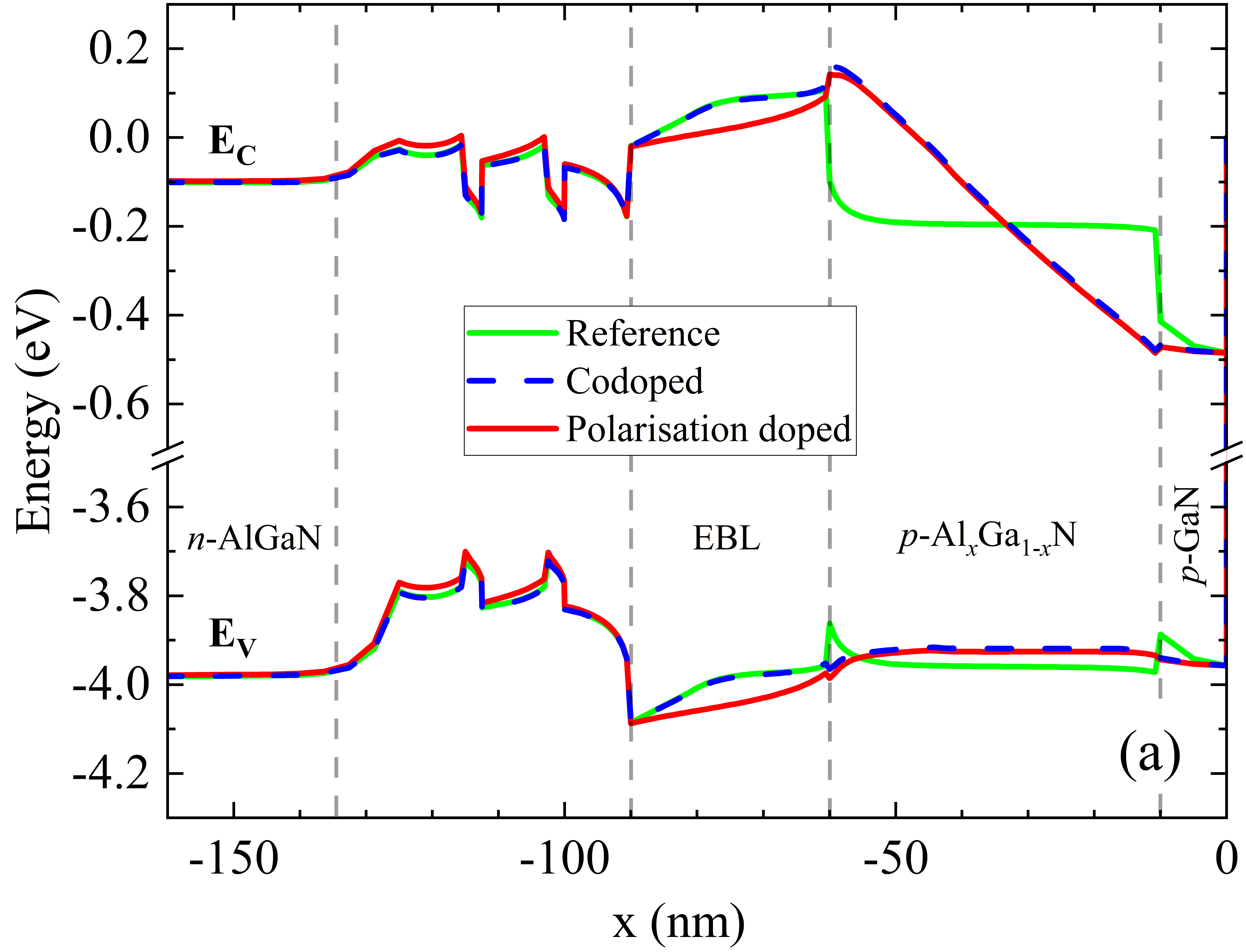}
    \phantomcaption
    \label{fig:fig3a}
    \end{subfigure}\\
    \begin{subfigure}[b]{0.5\textwidth}
    \centering
    \includegraphics[scale=0.40]{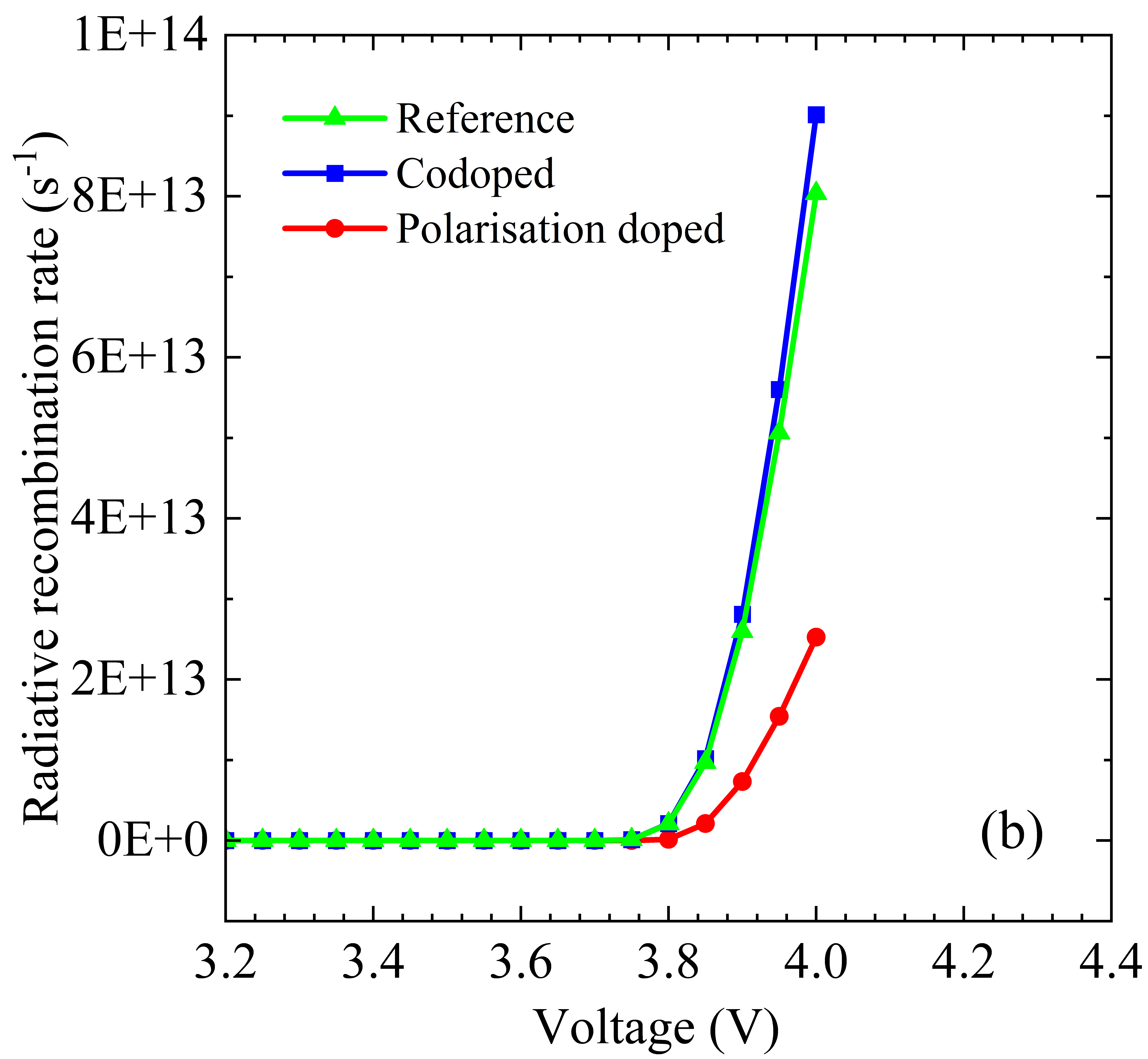}
    \phantomcaption
    \label{fig:fig3b}
    \end{subfigure}%
\caption{(a) Simulated band diagrams for the three LED structures at a bias of 3.9 V, and (b) the corresponding radiative recombination rates as a function of bias.}
\end{figure}

An improvement in EL intensity is observed in the co-doped LED compared to the reference LED, which might be attributable to the higher hole concentration in the co-doped structure allowing an improved device injection efficiency. To examine this further, numerical simulations were performed with the TCAD package Silvaco. The LED structures were simulated at room temperature, as per the \emph{p}-doping approaches in the three experimentally prepared LEDs. An acceptor concentration ($N_A$) of $1\times10^{18}$ cm$^{-3}$ was set in both the co-doped and reference \emph{p}-AlGaN regions, while a polarisation-doped hole concentration was generated by the graded Al layers. Figure \ref{fig:fig3a} shows the simulated band diagrams of the three structures at an applied bias of 3.9 V. The LEDs with graded \emph{p}-AlGaN layers show the band bending occurring mainly in the conduction band, to maintain the quasi-Fermi level position relative to the valence band. In the EBL region, the bands in the polarisation doped LED are reduced in energy by approximately 60 meV. The lack of Mg doping in the EBL forces the bands downward to accommodate a higher hole quasi-Fermi level. The resulting weaker barrier effect to electron overflow may partly explain the measured poorer EL performance from this LED \cite{chitnis2003improved}. Figure \ref{fig:fig3b} in fact shows a lower radiative recombination rate from this simulated LED, which can be attributed to the reduced injection efficiency. 

A further feature of the simulation is that grading the \emph{p}-Al$_x$Ga$_{1-x}$N region results in the elimination of the polarisation discontinuity at the \emph{p}-AlGaN/EBL interface, which prevents the formation of a two-dimensional hole gas (2DHG). This elimination enhances the hole injection into the active region \cite{schubert2018light}. However, it appears the EBL enhancement induced by the Mg doping is the dominant effect, as shown in Figure \ref{fig:fig3b}, and consistent with the experimental observation. These results indicate that the injection efficiency is a significant factor governing the performance benefits in polarisation doped LEDs.

\subsection{Temperature-dependent characterisation}
\begin{figure}
\centering
\includegraphics[scale=0.5]{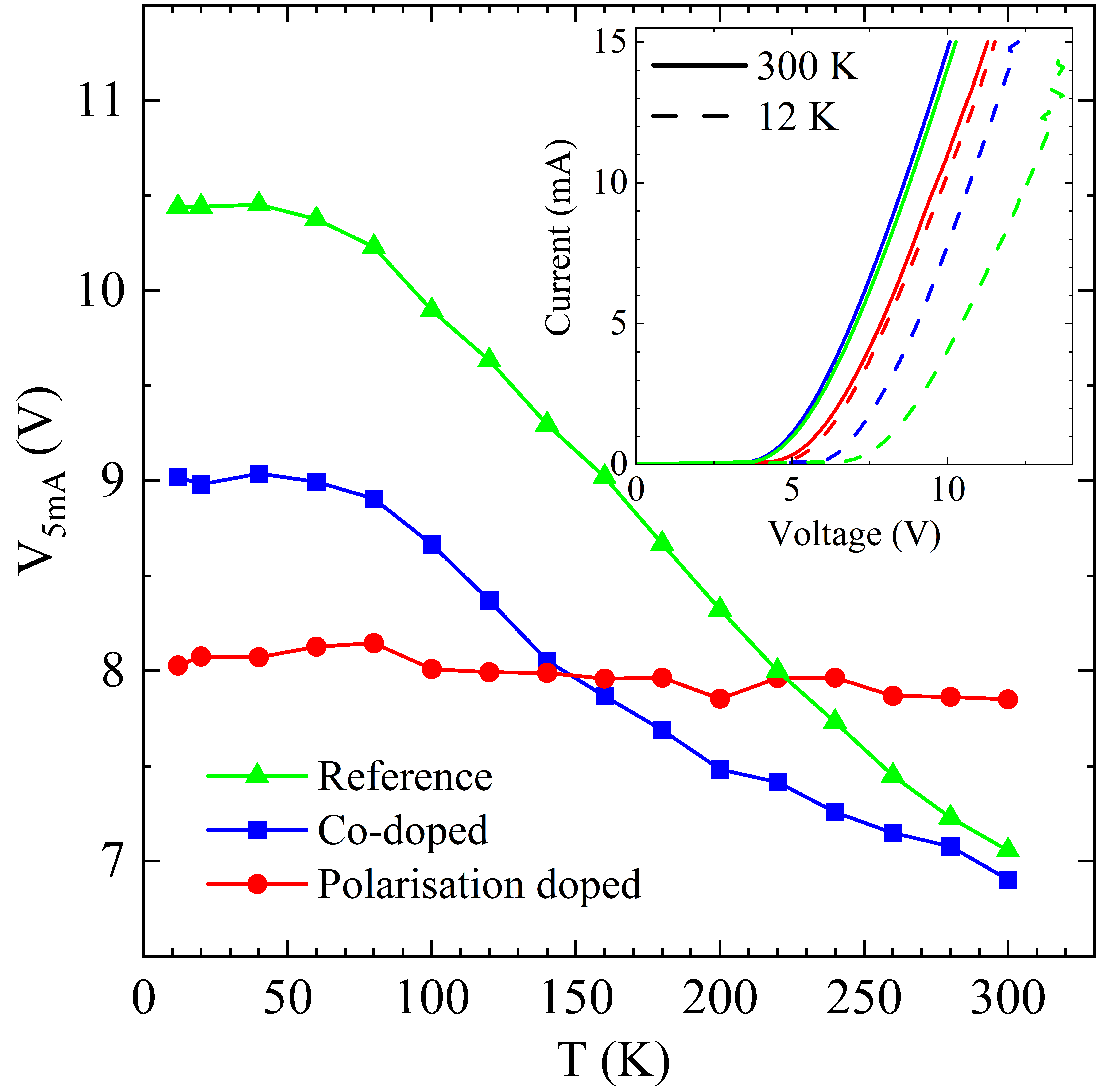}
\caption{\label{fig:fig4} Voltage at 5 mA as a function temperature. Inset are the IV characteristics at 300 K (solid) and 12 K (dashed).}
\end{figure}

Figure \ref{fig:fig4} shows the LED voltage at 5 mA as a function of temperature. At 300 K, there is no significant difference between the two LEDs with Mg doping, however, when the Mg is removed and only polarisation doping is relied upon, an increase of $\sim$1\,V is observed. At 12\,K, the voltages corresponding to 5 mA increase by $\sim$2\,V and $\sim$3\,V for the co-doped and reference LEDs, respectively, while the polarisation doped LED shows almost no increase. The contrasting dependence of voltage on temperature between the three LEDs reveals the insensitivity of the field-ionised hole concentration to thermal energy. 

There is no significant difference between the series resistances, $R_s$, of the LEDs, seen in the current-voltage (IV) characteristics inset in Figure \ref{fig:fig4}. This can be understood by considering the small contribution to $R_s$ from the thin \emph{p}-region (the lateral resistance of the \emph{n}-type current spreading layer likely dominates the overall $R_s$ of the devices). 

At low temperature, one might expect the co-doped LED to increase in operating voltage to the same value as the polarisation doped LED, as the hole concentrations in both should be created solely by field-ionisation. However, the voltage across the co-doped LED increases monotonically with decreasing temperature to 1 V higher than the polarisation doped LED. The reason for this behaviour is not clear, yet the presence of (non-activated) Mg impurity atoms may still have an impact on device performance. The energy barrier at the \emph{p}-GaN/\emph{p}-metal interface has been proposed to be a contributing factor to the significantly increased operating voltage at low temperature \cite{cao2003temperature}. However, whether this barrier is determined by the free hole concentration or the net $N_A - N_D$ concentration is not widely accepted. Degradation of the \emph{n}-contact at low temperatures has also been suggested as a cause \cite{chitnis2002low}, yet from the temperature-independent voltage of the polarisation doped LED, this is unlikely to be the case. 

\begin{figure*}
\centering
\includegraphics[scale=0.35]{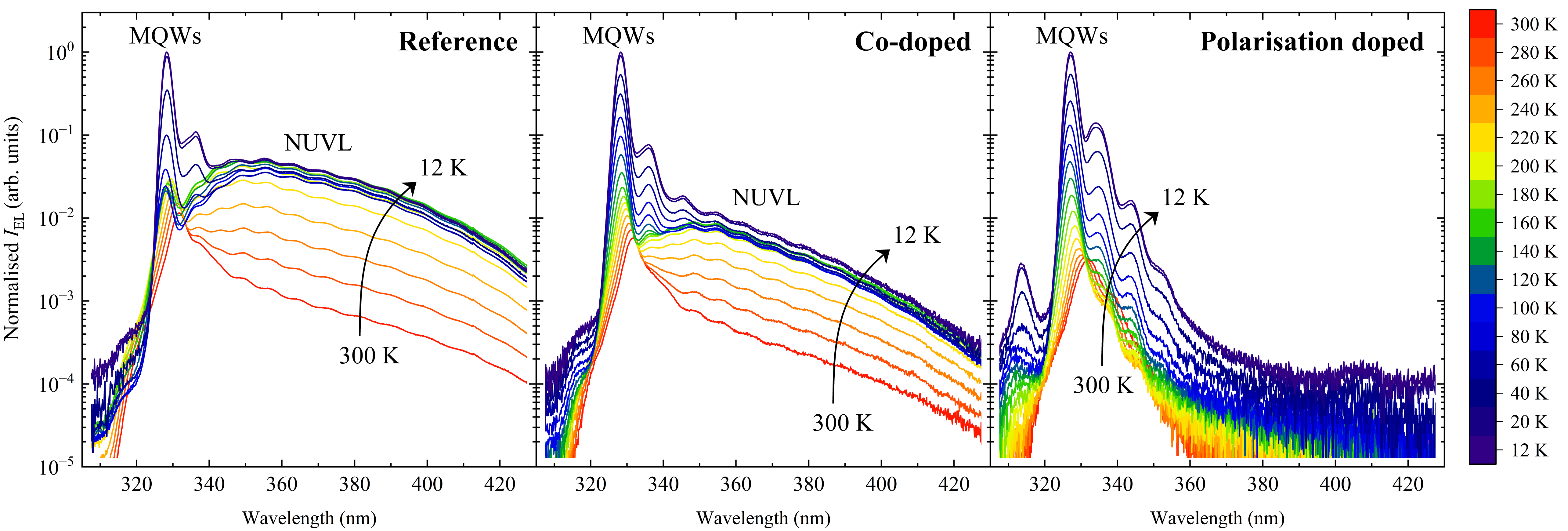}
\caption{\label{fig:fig5}Temperature-dependent EL spectra of the LEDs, measured at an injection current of 5 mA.}
\end{figure*}

\begin{figure}
\centering
\includegraphics[scale=0.5]{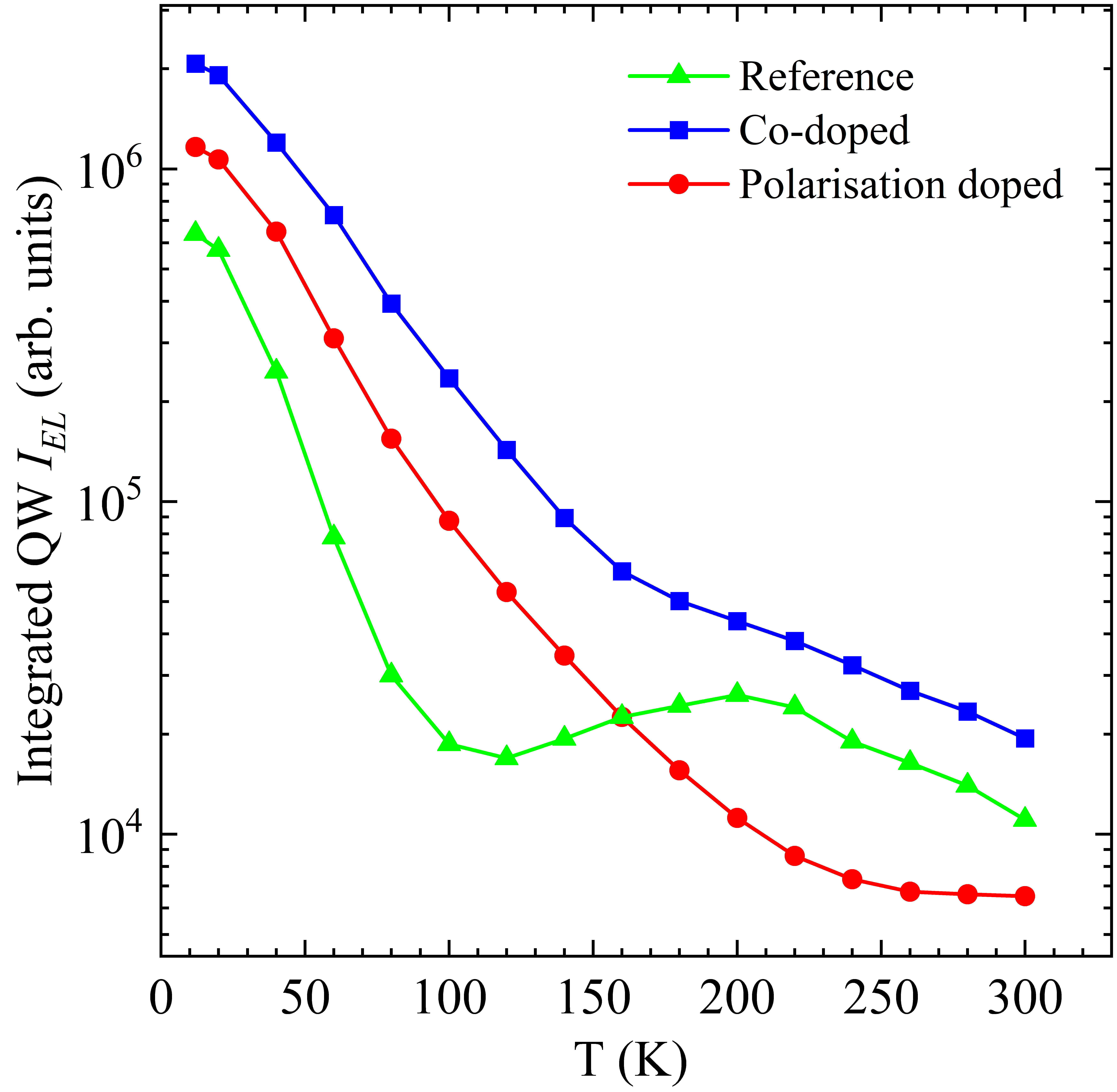}
\caption{\label{fig:fig6} Integrated QW EL intensity as a function of temperature, calibrated based on their relative intensities at 300 K.}
\end{figure}

The temperature-dependent EL spectra of the LEDs are plotted on a semi-logarithmic scale in Figure \ref{fig:fig5}, normalised to their intensities at 12 K. The injection current was 5\,mA (approx. 60\,A\,cm$^{-2}$) for all temperature-dependent measurements, which is well below any observed droop in output power potentially stemming from thermal effects. Figure \ref{fig:fig6} shows the integrated intensities of the respective QW peaks versus temperature, calibrated based on their relative intensities at 300 K. The temperature dependence of the EL intensity of the reference LED is rather complicated (not monotonic): it has a local maximum at $\sim$200\,K and a local minimum at $\sim$120\,K. The internal quantum efficiency (IQE) is expected to only rise from high to low temperature, which is confirmed by the temperature-resolved PL for this sample (not shown). As such, the EL intensity initially rises with reducing temperature. However, with no polarisation doping to support hole conductivity at lower temperatures, a strong deterioration of injection efficiency (IE) is reflected by the degradation of EL intensity from 200\,K to 120\,K. This deterioration is accompanied by a larger voltage increase (to keep current at 5\,mA).

The strong increase in EL intensity below 100\,K for the reference LED is in contradiction to the expectation that the thermally-ionised hole concentration will freeze-out at low temperatures \cite{lancefield2001temperature, moe2012impact}.  It is clear that in our case, it is caused by a strong rise in IQE which is expected, but still leaves a question as to the origin of the holes at these extremely low temperatures. This `S' shaped dependence of EL intensity on temperature has previously been observed by Lee \emph{et al}. \cite{lee2001temperature}, but no explanation or speculations have been provided by the authors for this effect. 

One possible origin of holes at the lowest temperatures (<\,100 K) is due to the large voltage drop across this LED, even though it is counter-intuitive at first glance. As thermally activated acceptors become frozen, fixed current excitation results in an increased voltage, for this LED especially (>\,10 V at \emph{T} <\,100\,K). The calculated effective field across the \emph{p}-type region at this bias is on the order of 1\,MV cm$^{-1}$, which is comparable to the field produced by polarisation doping \cite{zhang2011theoretical}. At such a high electric field, there is sufficient energy to start field-ionising the holes from the now-frozen Mg acceptors, thus supplying a hole concentration to the active region. This might be the reason why the voltage does not rise further (it saturates at 10.5\,V below 50\,K as seen in Figure \ref{fig:fig4}). As the V$_{5 mA}$ saturates at low temperature, so too does the IE. On the other hand, the IQE is expected to strongly increase in this temperature range (as seen also from both polarisation doped and co-doped LEDs). This apparent saturation in IE, coupled with the increasing IQE (towards 12\,K), can explain the sudden increase in EL intensity observed. 

The alternatives to the mechanism involving field-ionised holes (due to large bias) in the above explanation might be: the presence of a 2DHG at the EBL/\emph{p}-AlGaN interface (Figure \ref{fig:fig3a}); and nearest-neighbour hopping conduction between spatially adjacent Mg impurity atoms, previously reported for \emph{p}-(Al)GaN at low temperatures \cite{gotz1999hall, gunning2012negligible, kajikawa2017hall}. A detailed investigation into the carrier transport mechanisms is outside the scope of this report, however, further investigations are ongoing. 

As expected from the temperature-independent characteristic of a polarisation-doped carrier concentration, strong QW-related luminescence is observed at 12\,K in both of the LEDs with graded Al \emph{p}-type regions. From Figure \ref{fig:fig6}, both of these LEDs show a general monotonic decrease in intensity with increasing temperature, presenting similar quenching behaviour up to approximately 150\,K. At this temperature, the co-doped LED experiences a change in slope, indicating a decrease in quenching. The behaviour of the co-doped LED just below room temperature is similar to that of the reference LED, which indicates that the electrical properties in this temperature range are dominated by the Mg acceptors. Meanwhile, the polarisation doped LED continues an exponential decrease in intensity, until the temperature approaches 300\,K at which point this quenching saturates. 

The IE along with the IQE and the light extraction efficiency (LEE) determine the external quantum efficiency (EQE) of an LED, given by: 
\begin{equation} \label{eq:EQE}
EQE = IE \times IQE \times LEE.
\end{equation}
The LEE should remain approximately constant both across the temperature range and between LEDs. The integrated EL intensity in Figure \ref{fig:fig6} can therefore be used as a proxy for the trend in EQE with temperature. 
 
The TDD is approximately equal between all templates. From equation \ref{eq:EQE}, the temperature dependent behaviour of the polarisation doped LED should be mostly explainable in terms of IQE as its electrical properties are largely temperature-independent. The saturation in quenching towards 300 K would suggest that most of the defects that act as non-radiative recombination centres are already activated above $\sim$260\,K, and this seems to be the reason why no further EQE deterioration occurs above this temperature. Meanwhile, the presence of Mg can only have adverse or ideally no effects on IQE. Therefore, the slower EL quenching in the co-doped LED above 150\,K is likely related to the unfreezing of Mg acceptors and a corresponding increase in \emph{p}-type conductivity, which should contribute to an enhancement in IE. Indeed, the voltage at 5\,mA in the co-doped LED happens to drop below that of the polarisation doped LED exactly at 150\,K (Figure \ref{fig:fig4}).

\subsection{Analysis of NUVL vs T}
The rate at which electrons are captured by a trap state, $R_n$, can be expressed by:
\begin{equation}
R_n = C_nN_t[1 - f_F(E_t)]n,
\end{equation}
where $C_n$ is a constant proportional to the electron-capture cross section, $N_t$ is the concentration of trap states, $n$ is the concentration of electrons in the conduction band, and $f_F(E_t)$ is the probability that the trap contains an electron \cite{neamen2011semiconductor}. The rate at which carriers are captured by the Mg-related defects is therefore proportional to the number of electrons in this region. Considering the recombination mechanism associated with this defect state and assuming $N_t$ is roughly the same between LEDs due to nominally identical Mg concentrations, an increased number of electrons overflowing from the active region will result in an increased intensity of this NUVL band (inset Figure \ref{fig:fig2}).

\begin{figure}[hbt]
\centering
\includegraphics[scale=0.5]{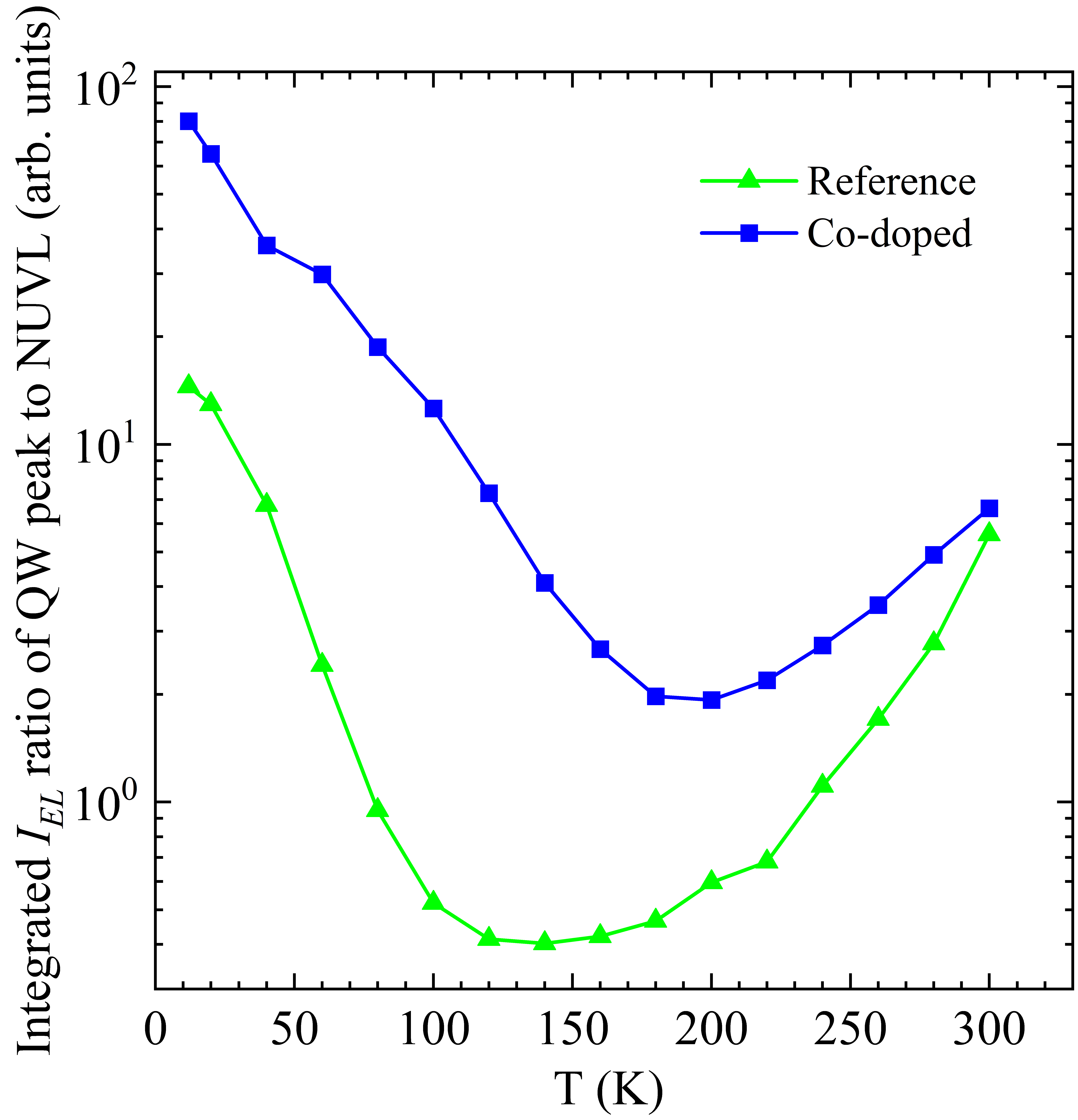}
\caption{\label{fig:Defect_summary} Ratio of QW peak and NUVL intensities. No significant NUVL is observed in the polarisation doped LED.}
\end{figure}

The evolution of the NUVL with temperature can also provide insight into the carrier dynamics. Figure \ref{fig:Defect_summary} shows the ratio of the QW peak intensities to the NUVL intensities. No such peak is observed in the polarisation doped LED, hence its absence from this analysis. As was mentioned earlier, the origin of this broad peak is attributed to overflow of electrons into the \emph{p}-type region and subsequent recombination involving Mg-related defect states \cite{kaufmann1998, eckey1998photoluminescence,reshchikov1999}. In both of the Mg-containing LEDs, when the temperature is reduced, the intensity of this band increases, until about 180\,K in the former and 120\,K in the latter. The fact that the ratio of the reference LED presents a much larger drop with decreasing temperature indicates that the hole freezeout, and as a result electron overflow out of the active region is more severe without the aid of polarisation doping as was already alluded to above. The NUVL band is directly associated with recombination in Mg-doped regions and therefore is proportional to electron overflow. It is important to note that the efficiency of the transition associated with the NUVL should also increase with decreasing temperature, which would also act to increase the NUVL intensity. Regardless, it is the relative behaviour between the two LEDs which supports the mechanism of enhanced IE resulting in the reduced rate of EQE quenching above 150\,K.

\begin{figure}[htb]
\centering
\includegraphics[scale=0.5]{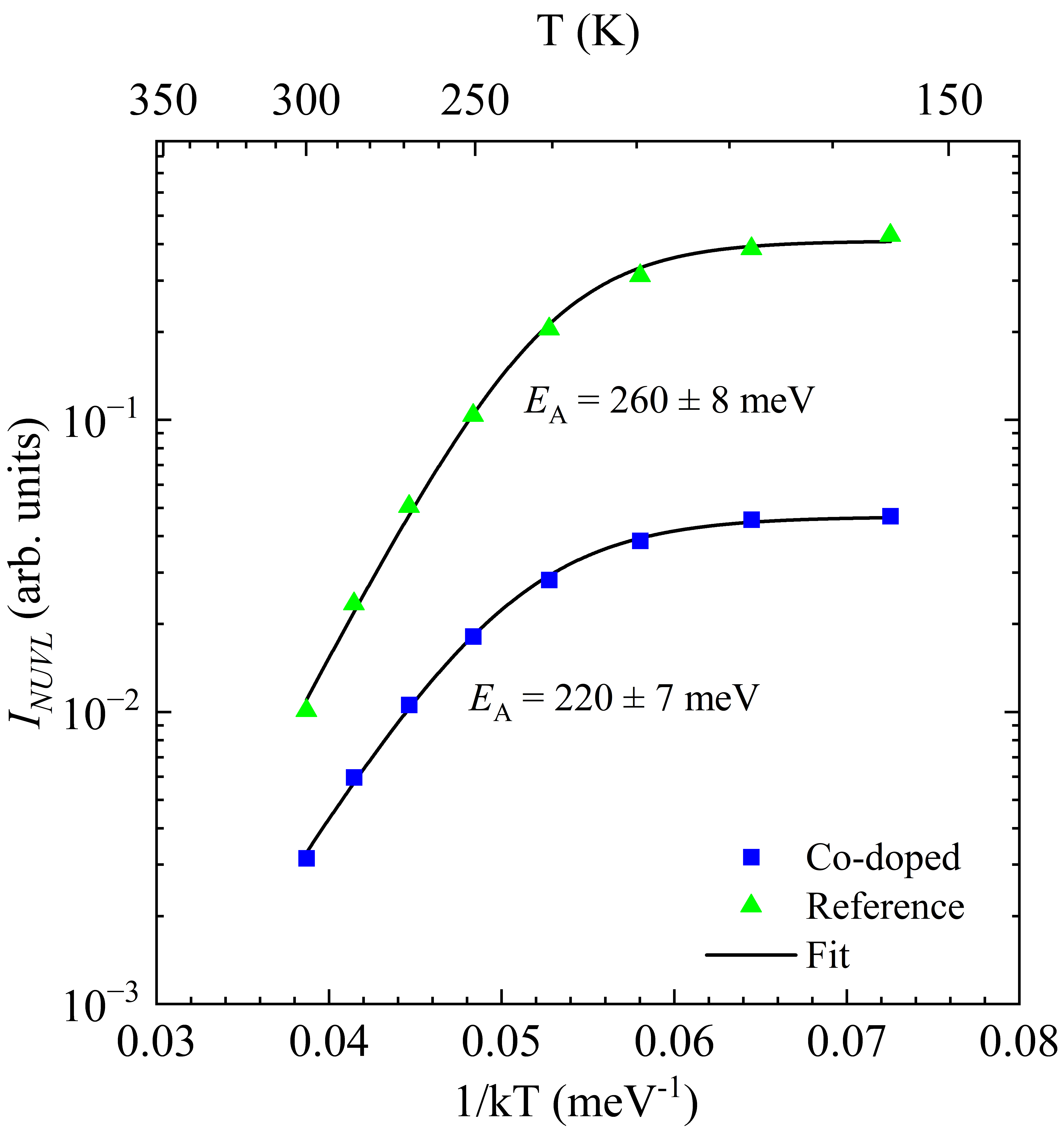}
\caption{\label{fig:defectEa} Arrhenius plot of the NUVL intensity for the two Mg-containing LEDs, using equation \ref{eq:LvsT}.}
\end{figure}

To further investigate this, an Arrhenius plot of the integrated intensity of NUVL was plotted in Figure \ref{fig:defectEa}. The intensity, $L$, was fitted with the equation for a single quenching process with an activation energy $E_A$:
\begin{equation} \label{eq:LvsT}
L = \frac{L_0}{1 + A~\mathrm{exp}(-E_A/kT)},
\end{equation}
where $L_0$ is the light intensity extrapolated to 0\,K, $A$ is a constant and $k$ is Boltzmann's constant. Activation energies of 220\,meV and 260\,meV were extracted for the co-doped and reference LEDs, respectively. 

One interpretation of the fitted energy values could be as follows. When the temperature is reduced from 300\,K, the number of inactivated Mg atoms increases due to its large activation energy. This increased concentration of non-ionised acceptor states leads to a rise in the transition between these acceptor states to deep-level donor point defects (and also conduction band edge), causing the initial increase in NUVL intensity. Thus the value of $E_A$ associated with the reduced intensity of this peak can be interpreted as an insight into the activation energy of the Mg dopant, which is comparable to values reported in the literature for this composition range \cite{suzuki1998doping, li2002}. Therefore, the lower value of $E_A$ in the LED with the graded Al layer compared to the LED with a uniform composition \emph{p}-region indicates an increase in the activation efficiency of Mg. This can be understood by considering the effect of compositional grading on the valence band. An enhanced activation of Mg could partly explain the increased EL intensity in the co-doped LED, and the improved ratio of QW-related luminescence to NUVL. 

It is again worth noting here that the behaviour of this NUVL as a function of temperature may also be affected by its varying efficiency as the recombination lifetimes associated with this transition change. It is important to focus here on the relative difference between the two Mg-containing LEDs, and not interpret this as a direct method for determining the dopant activation energy. 

Ultimately, the device performance becomes less sensitive to temperature with the introduction of polarisation doping, as well as increased QW-related EL intensity across the temperature range. While Sato \emph{et al.} \cite{sato2021} propose that for \emph{p}-type polarisation doping at higher Al compositions, the higher activation energy of Mg leads to more compensating point defects which compensate the polarisation doped hole concentration, the results presented here suggest that for lower Al compositions, Mg doping is beneficial alongside polarisation doping. 

Finally, we discuss the weak emission at shorter wavelengths compared to the main QW-related peak. It appears at lowest temperatures as only a shoulder in the two Mg-doped LEDs, and it is visible as a distinct peak in the polarisation doped LED. Similar emission has already been observed in AlGaN-based UV LEDs by De Santi \emph{et al.} \cite{de2017recombination}. We attribute the origin of this peak to radiative recombination in the QBs. At low temperatures, the thermal energy is not enough to delocalise the carriers which happen to be trapped in shallow localised states in the barriers. On top of this, non-radiative centres also become inactive at low temperature. This explains why, in principle, this peak becomes visible at low temperature. Regarding its relatively increased intensity in the polarisation doped LED, we speculate that the suppression of this peak in the Mg-containing LEDs is related to Mg back-diffusion \cite{kohler2005control}, allowing non-radiative recombination centres to suppress the luminescence even at low temperature. This suggests that this luminescence might be specifically from the final barrier, which experiences a large band-bending at the barrier/EBL interface, forming a quasi-well according to theoretical calculations using the classical drift-diffusion model \cite{zhang2011theoretical}.

\section{Conclusion}
In summary, 330\,nm UV LEDs with varied \emph{p}-doping structures have been prepared, and their electrical and luminescence behaviour studied as a function of temperature. For room temperature operation, it was found that an LED structure with both polarisation doping and Mg doping produces a higher EL intensity. Meanwhile, Hall effect measurements of corresponding test structures also show that the highest hole concentration was achieved with a co-doping strategy. In the LED with only polarisation doping and no Mg, a cleaner spectrum was achieved due to the elimination of NUVL stemming from Mg-related defect states in the \emph{p}-region. 

The expected complete freeze-out of the holes at low temperature in the LED with only Mg doping was not observed, although it demonstrated a significantly increased forward bias to maintain fixed current injection. There was a complex progression of EL intensity with temperature, which may be due to an interplay between increasing IQE, decreasing IE, and another yet-unidentified mechanism of hole conductivity at low temperatures. Meanwhile, both LEDs with polarisation doping showed a steady increase in EL intensity when the temperature was decreased from 300\,K to 12\,K, as expected from the increasing IQE along with the temperature-independent field-ionised hole concentration.  

The overall improvement in performance with polarisation doping is attributed to both reduced electron overflow and increased hole injection, resulting in an improved IE, which is supported by simulation. Rate equation analysis of the NUVL dependence on temperature suggests an enhanced Mg activation efficiency also contributes to the improved performance. Therefore, for lower Al compositions, we propose that (\emph{p}-type) polarisation doping benefits from co-doping with Mg. Further optimisation of the doping conditions could allow for increased EL intensity as well as a cleaner emission spectrum, realising higher power devices without unwanted parasitic emission.

\ack 
The authors gratefully acknowledge the funding from Research Ireland under grant no. 12/RC/2276\_P2.

\section*{Data Availability Statement}

The data that supports the findings of this study are available upon reasonable request from the authors.


\section*{References}
\bibliographystyle{unsrt}
\bibliography{myBib}

\begin{thebibliography}{10}

\bibitem{kneissl2019}
Michael Kneissl, Tae-Yeon Seong, Jung Han, and Hiroshi Amano.
\newblock {The emergence and prospects of deep-ultraviolet light-emitting diode technologies}.
\newblock {\em Nature Photonics}, 13(4):233--244, 2019.

\bibitem{amano2020}
Hiroshi Amano, Ram{\'o}n Collazo, Carlo De~Santi, Sven Einfeldt, Mitsuru Funato, Johannes Glaab, Sylvia Hagedorn, Akira Hirano, Hideki Hirayama, Ryota Ishii, et~al.
\newblock {The 2020 UV emitter roadmap}.
\newblock {\em Journal of Physics D: Applied Physics}, 53(50):503001, 2020.

\bibitem{imura2007}
M~Imura, N~Kato, N~Okada, K~Balakrishnan, M~Iwaya, S~Kamiyama, H~Amano, I~Akasaki, T~Noro, T~Takagi, et~al.
\newblock {Mg‐doped high‐quality AlxGa1–xN (x= 0‐1) grown by high‐temperature metal‐organic vapor phase epitaxy}.
\newblock {\em physica status solidi c}, 4(7):2502--2505, 2007.

\bibitem{li2002}
J~Li, TN~Oder, ML~Nakarmi, JY~Lin, and HX~Jiang.
\newblock {Optical and electrical properties of Mg-doped p-type Al{$_x$}Ga{$_{1-x}$}N}.
\newblock {\em Applied physics letters}, 80(7):1210--1212, 2002.

\bibitem{zhang2022}
Ziyi Zhang, Maki Kushimoto, Akira Yoshikawa, Koji Aoto, Chiaki Sasaoka, Leo~J Schowalter, and Hiroshi Amano.
\newblock {Continuous-wave lasing of AlGaN-based ultraviolet laser diode at 274.8 nm by current injection}.
\newblock {\em Applied Physics Letters}, 121(22), 2022.

\bibitem{cao2023}
Yi-Wei Cao, Quan-Jiang Lv, Tian-Peng Yang, Ting-Ting Mi, Xiao-Wen Wang, Wei Liu, and Jun-Lin Liu.
\newblock {Realization of high-efficiency AlGaN deep ultraviolet light-emitting diodes with polarization-induced doping of the p-AlGaN hole injection layer}.
\newblock {\em Chinese Physics B}, 32(5):058503, 2023.

\bibitem{kolbe2023}
Tim Kolbe, Arne Knauer, Jens Rass, Hyun~Kyong Cho, Sylvia Hagedorn, Fedir Bilchenko, Anton Muhin, Jan Ruschel, Michael Kneissl, Sven Einfeldt, et~al.
\newblock {234 nm far-ultraviolet-C light-emitting diodes with polarization-doped hole injection layer}.
\newblock {\em Applied Physics Letters}, 122(19), 2023.

\bibitem{simon2010}
John Simon, Vladimir Protasenko, Chuanxin Lian, Huili Xing, and Debdeep Jena.
\newblock {Polarization-induced hole doping in wide–band-gap uniaxial semiconductor heterostructures}.
\newblock {\em Science}, 327(5961):60--64, 2010.

\bibitem{sato2021}
Kosuke Sato, Kazuki Yamada, Konrad Sakowski, Motoaki Iwaya, Tetsuya Takeuchi, Satoshi Kamiyama, Yoshihiro Kangawa, Pawel Kempisty, Stanislaw Krukowski, Jacek Piechota, et~al.
\newblock {Effects of Mg dopant in Al-composition-graded Al{$_x$}Ga{$_{1-x}$}N (0.45{$\le$}x) on vertical electrical conductivity of ultrawide bandgap AlGaN p–n junction}.
\newblock {\em Applied Physics Express}, 14(9):096503, 2021.

\bibitem{van1999doping}
Chris~G Van~de Walle, Catherine Stampfl, J{\"o}rg Neugebauer, MD~McCluskey, and NM~Johnson.
\newblock {Doping of AlGaN alloys}.
\newblock {\em Materials Research Society Internet Journal of Nitride Semiconductor Research}, 4(S1):890--901, 1999.

\bibitem{ishii2020temperature}
Ryota Ishii, Akira Yoshikawa, Kazuhiro Nagase, Mitsuru Funato, and Yoichi Kawakami.
\newblock {Temperature-dependent electroluminescence study on 265-nm AlGaN-based deep-ultraviolet light-emitting diodes grown on AlN substrates}.
\newblock {\em AIP Advances}, 10(12), 2020.

\bibitem{chitnis2002low}
A~Chitnis, R~Pachipulusu, V~Mandavilli, M~Shatalov, E~Kuokstis, JP~Zhang, V~Adivarahan, S~Wu, Grigory Simin, and M~Asif Khan.
\newblock {Low-temperature operation of AlGaN single-quantum-well light-emitting diodes with deep ultraviolet emission at 285 nm}.
\newblock {\em Applied Physics Letters}, 81(16):2968, 2002.

\bibitem{grzanka2007role}
S~Grzanka, G~Franssen, G~Targowski, K~Krowicki, T~Suski, R~Czernecki, P~Perlin, and M~Leszczy{\'n}ski.
\newblock {Role of the electron blocking layer in the low-temperature collapse of electroluminescence in nitride light-emitting diodes}.
\newblock {\em Applied Physics Letters}, 90(10), 2007.

\bibitem{zhang2010low}
JC~Zhang, Y~Sakai, and T~Egawa.
\newblock {Low-temperature electroluminescence quenching of AlGaN deep ultraviolet light-emitting diodes}.
\newblock {\em Applied Physics Letters}, 96(1), 2010.

\bibitem{yasuda2017relationship}
Toshiki Yasuda, Tetsuya Takeuchi, Motoaki Iwaya, Satoshi Kamiyama, Isamu Akasaki, and Hiroshi Amano.
\newblock {Relationship between lattice relaxation and electrical properties in polarization doping of graded AlGaN with high AlN mole fraction on AlGaN template}.
\newblock {\em Applied Physics Express}, 10(2):025502, 2017.

\bibitem{kaufmann1998}
U~Kaufmann, M~Kunzer, M~Maier, H~Obloh, A~Ramakrishnan, B~Santic, and P~Schlotter.
\newblock {Nature of the 2.8 eV photoluminescence band in Mg doped GaN}.
\newblock {\em Applied physics letters}, 72(11):1326--1328, 1998.

\bibitem{chitnis2003improved}
A~Chitnis, JP~Zhang, V~Adivarahan, M~Shatalov, S~Wu, R~Pachipulusu, V~Mandavilli, and M~Asif Khan.
\newblock {Improved performance of 325-nm emission AlGaN ultraviolet light-emitting diodes}.
\newblock {\em Applied physics letters}, 82(16):2565--2567, 2003.

\bibitem{schubert2018light}
E~Fred Schubert.
\newblock {\em {Light-Emitting Diodes (2018)}}.
\newblock E. Fred Schubert, 2018.

\bibitem{cao2003temperature}
XA~Cao, SF~Leboeuf, LB~Rowland, and H~Liu.
\newblock {Temperature-dependent electroluminescence in InGaN/GaN multiple-quantum-well light-emitting diodes}.
\newblock {\em Journal of electronic materials}, 32:316--321, 2003.

\bibitem{lancefield2001temperature}
D~Lancefield and H~Eshghi.
\newblock {Temperature-dependent hole transport in GaN}.
\newblock {\em Journal of Physics: Condensed Matter}, 13(40):8939, 2001.

\bibitem{moe2012impact}
Craig~G Moe, Gregory~A Garrett, Paul Rotella, Hongen Shen, Michael Wraback, Max Shatalov, Wenhong Sun, Jianyu Deng, Xuhong Hu, Yuri Bilenko, et~al.
\newblock {Doping characteristics and electrical properties of Mg-doped AlGaN grown by atmospheric-pressure MOCVD}.
\newblock {\em Applied Physics Letters}, 101(25), 2012.

\bibitem{lee2001temperature}
Chia-Ming Lee, Chang-Cheng Chuo, Jing-Fu Dai, Xian-Fa Zheng, and Jen-Inn Chyi.
\newblock {Temperature dependence of the radiative recombination zone in InGaN/GaN multiple quantum well light-emitting diodes}.
\newblock {\em Journal of Applied Physics}, 89(11):6554--6556, 2001.

\bibitem{zhang2011theoretical}
Lian Zhang, K~Ding, NX~Liu, TB~Wei, XL~Ji, P~Ma, JC~Yan, JX~Wang, YP~Zeng, and JM~Li.
\newblock {Theoretical study of polarization-doped GaN-based light-emitting diodes}.
\newblock {\em Applied Physics Letters}, 98(10), 2011.

\bibitem{gotz1999hall}
W~G{\"o}tz, RS~Kern, CH~Chen, H~Liu, DA~Steigerwald, and RM~Fletcher.
\newblock {Hall-effect characterization of III--V nitride semiconductors for high efficiency light emitting diodes}.
\newblock {\em Materials Science and Engineering: B}, 59(1-3):211--217, 1999.

\bibitem{gunning2012negligible}
Brendan Gunning, Jonathan Lowder, Michael Moseley, and W~Alan~Doolittle.
\newblock {Negligible carrier freeze-out facilitated by impurity band conduction in highly p-type GaN}.
\newblock {\em Applied Physics Letters}, 101(8), 2012.

\bibitem{kajikawa2017hall}
Yasutomo Kajikawa.
\newblock {Hall factor for hopping conduction in n-and p-type GaN}.
\newblock {\em physica status solidi c}, 14(1-2):1600129, 2017.

\bibitem{neamen2011semiconductor}
Donald~A Neamen and Dhrubes Biswas.
\newblock {\em {Semiconductor physics and devices}}.
\newblock McGraw-Hill higher education New York, 2011.

\bibitem{eckey1998photoluminescence}
L~Eckey, U~Von~Gfug, J~Holst, A~Hoffmann, A~Kaschner, H~Siegle, C~Thomsen, B~Schineller, K~Heime, M~Heuken, et~al.
\newblock {Photoluminescence and Raman study of compensation effects in Mg-doped GaN epilayers}.
\newblock {\em Journal of applied physics}, 84(10):5828--5830, 1998.

\bibitem{reshchikov1999}
MA~Reshchikov, G-C Yi, and BW~Wessels.
\newblock {Behavior of 2.8-and 3.2-eV photoluminescence bands in Mg-doped GaN at different temperatures and excitation densities}.
\newblock {\em Physical Review B}, 59(20):13176, 1999.

\bibitem{suzuki1998doping}
Mariko Suzuki, Johji Nishio, Masaaki Onomura, and Chie Hongo.
\newblock {Doping characteristics and electrical properties of Mg-doped AlGaN grown by atmospheric-pressure MOCVD}.
\newblock {\em Journal of crystal growth}, 189:511--515, 1998.

\bibitem{de2017recombination}
Carlo De~Santi, Matteo Meneghini, Desiree Monti, Johannes Glaab, Martin Guttmann, Jens Rass, Sven Einfeldt, Frank Mehnke, Johannes Enslin, Tim Wernicke, et~al.
\newblock {Recombination mechanisms and thermal droop in AlGaN-based UV-B LEDs}.
\newblock {\em Photonics Research}, 5(2):A44--A51, 2017.

\bibitem{kohler2005control}
K~K{\"o}hler, T~Stephan, A~Perona, J~Wiegert, M~Maier, M~Kunzer, and J~Wagner.
\newblock {Control of the Mg doping profile in III-N light-emitting diodes and its effect on the electroluminescence efficiency}.
\newblock {\em Journal of applied physics}, 97(10), 2005.

\end{thebibliography}

\end{document}